\documentclass[fleqn,10pt]{wlscirep}

\newtheorem{lemma}{{\bf Lemma}}

\usepackage{graphicx}
\usepackage{amssymb}
\usepackage{subcaption}
\usepackage{pgf,tikz}
\usepackage{pgfplots}
\usetikzlibrary{spy}
\usetikzlibrary{backgrounds}
\usetikzlibrary{decorations}
\makeatletter
\newcommand*{\rightharpoonupfill@}{%
  \arrowfill@\relbar\relbar\rightharpoonup
}

\newcommand*{\leftharpoondownfill@}{%
  \arrowfill@\leftharpoondown\relbar\relbar
}

\newcommand{\xrightleftharpoons}[2][]{%
  \ensuremath{%
    \mathrel{%
      \settoheight{\dimen@}{\raise 2pt\hbox{$\rightharpoonup$}}%
      \setlength{\dimen@}{-\dimen@}%
      \edef\CA@temp{\the\dimen@}%
      \settoheight\dimen@{$\rightleftharpoons$}%
      \addtolength{\dimen@}{\CA@temp}%
      \raisebox{\dimen@}{%
        \rlap{%
          \raisebox{2pt}{%
            $%
            \ext@arrow 0359\rightharpoonupfill@{\hphantom{#1}}{#2}%
            $%
          }%
        }%
        \hbox{%
          $%
          \ext@arrow 3095\leftharpoondownfill@{#1}{\hphantom{#2}}%
          $%
        }%
      }%
    }%
  }%
}
\makeatother

\def\Div{\hbox{{\rm I}\kern-.2em\hbox{\rm Div}}}

\title{Analysis of biochemical mechanisms provoking differential spatial expression in Hh target genes}
\author[1,+]{Manuel Camb\'on}
\author[1,*,+]{\'Oscar  S\'anchez}
\affil[1]{University of Granada, Applied Mathematics Department,  Granada, E18071, Spain}
\affil[*]{ossanche@ugr.es}
\affil[+]{these authors contributed equally to this work}

\begin{abstract}
This work seeks to analyse  the transcriptional effects of some biochemical mechanisms 
 proposed in previous literature which attempts to explain the differential spatial expression of   Hedgehog target genes involved in \textit{Drosophila} development. 
Specifically, the expression of \textit{decapentaplegic} and \textit{patched}, genes whose transcription is believed to be controlled by the activator and repressor forms of the transcription factor Cubitus interruptus (Ci). 
This study is based on a thermodynamic approach which provides 
binding equilibrium weighted average rate expressions for genes controlled by transcription factors competing and 
(possibly) cooperating for common binding sites,
 in the same way that Ci's activator and repressor forms might do. 
These expressions are refined to produce simpler equivalent formulae allowing  their 
mathematical analysis.
 Thanks to this, we can evaluate the correlation between several 
molecular processes  and   biological features observed at tissular level.
In particular, we will focus on how  high/low/differential affinity and null/total/partial cooperation
modify the activation/repression regions of the target genes 
or provoke signal modulation.
\end{abstract}

\begin{document}
\flushbottom
\maketitle
\thispagestyle{empty}



\section*{Introduction}
{ Hedgehog (Hh)} is 
a morphogen,  signaling protein  acting on cells directly to  induce distinct cellular responses. It is involved in several developmental systems such as the \textit{Drosophila melanogaster} embryo.
In particular, in \textit{Drosophila} wing imaginal disc, the secretion of  Hh from cells in the posterior compartment  induces the expression of several target genes 
in the anterior compartment cells causing the patterning of the central domain of the wing~\cite{Tabata2004,Torroja2005}.
Cubitus interruptus (Ci), a { \textit{Drosophila}} transcription factor (TF), controls  the synthesis of the Hh target genes \textit{decapentaplegic}  ({\textit{dpp}}) and \textit{patched} ({\textit{ptc}}). 
It has been previously proposed that these genes  are regulated by sets of a promoter and $3$ binding sites (cis-regulatory elements, also known as enhancers), and TFs of the Cubitus interruptus family in two opposite forms 
competing for the same genomic binding sites
~\cite{Parker2011}. 
In the absence of Hh, Ci is cleaved to become a transcriptional repressor, CiR, but in the presence of Hh it is converted to the activator form, CiA.
However, the same Hh signaling causes differential spatial expression of these target genes: {\textit{ptc}} expression is 
restricted to the region of highest Hh signal concentration while {\textit{dpp}} responds more broadly in a lower zone of Hh signaling.

In the work of Parker and coauthors~\cite{Parker2011} the explanation of this fact was made considering several biochemical mechanisms in the binding of the TFs such as high-low, differential affinities and cooperativity. 
This discussion employs the fitting of a thermodynamic model  to discriminate between the different options. 
In~\cite{Barolo2013}, the same  discussion is focussed mainly on the high-low affinity character of the enhancers from a experimental point of view, concluding that low-affinity binding sites are required
 for normal {\textit{dpp}} activation in regions of relatively low signal. Another very interesting point in experiments shown in~\cite{Parker2011,Barolo2013} is the reinforcement  of the Hh signaling when 
 \textit{dpp} low-affinity Ci/Gli sites are converted to high-affinity sites, in the sense that both repression  and activation are generally stronger in the regions of respective net repression/activation.

 Here, we will reconsider the same questions performing the mathematical analysis of our own model  
building on the statistical thermodynamic method proposed 
by Shea, Ackers and coworkers \cite{Ackers1982,Shea1985}, also known as the BEWARE method\cite{Gilman2002}.
BEWARE (Binding Equilibrium Weighted Average Rate Expression)  is a well recognised method used frequently in mathematical modelling of gene transcription processes (see for instance \cite{Ay2011} for a general discussion). 
These weighted averages  give rise to lengthy mathematical expressions even for the case of   only two TFs. This impedes the possibility of deciphering biological effects without the use of numerical tools\cite{Zhou2008}. In addition, the averages involve a great amount of constants of diverse nature. Making approximation of these constants  constitutes a dilemma in itself, and the  effect of their modification in the biological system only has been tested by numerical and in vivo/vitro experiments  (see for instance \cite{Parker2011} or \cite{Junker2014}). 
Explicit simple analytical expressions have only been  proposed in the literature  for specific independent binding sites~\cite{Bintu2005}. 

In this work we formulated useful  easily applicable expressions for the BEWARE operator in the Hh signaling pathway where the pair of transcription factors CiA and CiR,  acting as activators and repressors, compete cooperatively for common enhancers. Based on the experiments done in the literature, we apply these expressions performing a mathematical analysis of the behaviour of the system under variations of the biochemical mechanisms considered: binding affinity, interaction intensity and cooperativity.  
These expressions allow us to  predict the effects of these mechanisms in presence of opposing activator/repressor TF gradients acting through the same cis-regulatory sites, a point not satisfactorily explained in any previous system\cite{Barolo2013}.

\section*{Results}
Concise expressions of the BEWARE operator have been deduced for a family of two TFs competing for common binding sites controlling transcription by the recruitment mechanism~\cite{Ptashne2005}
taking into consideration the following variables:
number of binding sites,
particular binding affinities  for each TF species (activators-repressors), 
posible effects of cooperativity between both TFs (total) or between the TFs of the same family (partial),
and finally interaction intensity for activators and repressors.
This operator determines the balance between opposing TFs that gives rise to equal gene activation rates. In particular, the mathematical analysis performed to this model predicts:
\begin{enumerate}
\item[i)] The existence of  relations between the concentrations of the transcription factors determining a threshold with respect to the activation basal level (meanly, transcription rates in absence of TFs). This threshold defines two areas: activated/repressed region for activator and repressor concentrations inducing transcription rates greater/lower than the basal.
This threshold is linear in the case of null or total cooperativity and involves more entangled expressions in the case of  partial cooperativity.
\item[ii)] The dependence of this threshold, and in consequence of the activated/repressed regions, on 
the relative affinity between activators and repressors, 
TFs interaction intensity, 
and partial cooperation.
\item[iii)]  Variations of the intensity of the signal due to proportional changes in the TFs affinity constants or total cooperativity, where we will refer to straightened signaling effects to transcription rate increments/decrements in the activated/repressed regions.
However, we remark that these effects do not change the activation/repression threshold.
\end{enumerate}
We propose that the differential response of two genes in the same cell containing the same TFs concentrations could be justified by the fact that the activation/repression regions are different for those genes. In consequence, the model proposes that the proportional (in particular, equal) low-high affinity of the TFs, or total cooperativity can not explain solely the differential spatial expression of {\textit{dpp}} and {\textit{ptc}} although they justify perfectly the stronger activation/repression measured in the literature~\cite{Parker2011,Barolo2013}  for higher affinity modified binding sites.
\section*{Methods:} \label{bifunctional}

\subsection*{Deduction of the  BEWARE operator}

As a first step, we will apply the ideas of the statistical thermodynamic method 
to 
the genes {\textit{dpp}} and {\textit{ptc}}, controlled by  the
transcription factors 
$\{CiA , CiR\}$. 
Thus, our goal in this point  is to deduce expressions for the time evolution of the concentration of protein $P$ (either be Dpp or Ptc) in terms of the concentrations of the TFs, i.e.,
\begin{equation}
\frac{d [P]}{dt} = \mbox{BEWARE}([CiA],[CiR])  \label{dP/dt_bifunctional}
\end{equation}
where `BEWARE()' represents a mathematical function specifying the dependence with respect to the activation/repression role of the TFs,  independently of other possible factors relevant for the protein evolution  as for instance degradation or space dispersion. In the model, the binding reactions of TFs and RNA polymerase (RNAP) in the enhancers 
and promoter, respectively, are much more faster than the synthesis of the protein P, hence it will be considered in thermodynamic equilibrium given by the Law of Mass Action. If $B$ is a set of non occupied enhancers-promoter, the complexes $BCiA$, $BCiR$ and $BRNAP$ have concentration at equilibrium given by
\begin{equation*}
[BCiA] = \frac{k_{+A}^{(1)}}{k_{-A}^{(1)}} [CiA] [B] := \frac{[CiA]}{K_A^{(1)}} [B]\, ,
\quad
[BCiR] = \frac{k_{+R}^{(1)}}{k_{-R}^{(1)}} [CiR] [B] := \frac{[CiR]}{K_R^{(1)}} [B]\, ,
\quad
[BRNAP] = \frac{k_{+RP}}{k_{-RP}} [RNAP] [B] := \frac{[RNAP]}{K_{RP}} [B]\, ,
\end{equation*}
where  $K_A^{(1)}$,  $K_R^{(1)}$ and $K_{RP}$ are  dissociation constants of the activators, repressors and RNA polymerase, so the quotients $\frac{[CiA]}{K_A^{(1)}}$, $\frac{[CiR]}{K_R^{(1)}}$ and $\frac{[RNAP]}{K_{RP}}$ are  dimensionless. The superscript $(1)$ stands for the dissociation constant of a reaction that takes place in absence of another TF previously bound in other enhancer (note that, since the sets only have one promoter, the superscript is not needed for the RNAP dissociation constant). The consecutive binding of more that one transcription factor is considered as a sequential and competitive process, such that the reactions

$$
CiA+ BCiA \xrightleftharpoons [k_{-A}^{(2)}]{k_{+A}^{(2)}} BCiACiA 
\quad \mbox{ or } \quad
CiR+ BCiA \xrightleftharpoons [k_{-R}^{(2)}]{k_{+R}^{(2)}} BCiACiR 
$$
will be given by equilibrium concentrations
$$[BCiACiA] = \frac{[CiA] [CiA] }{K_A^{(1)} K_A^{(2)}} [B] \quad \mbox{ and } \quad  [BCiACiR] = \frac{[CiA] [CiR] }{K_A^{(1)} K_R^{(2)}} [B]\, , $$
where now the superscript ${(2)}$ denotes the dissociation constant for a reaction of a TF that binds the operator with already one TF in some other site. On the other hand, the competition is modelled such that the dissociation constant of the free sites configuration does not depend on their position, but might depend on other existing bound TFs in the same set of enhancers by cooperativity or anticooperativity. 

We will denote by non cooperative TFs to all those proteins whose enhancer's affinity is not modified by any previously bound TFs, that is, they verify $K_A^{(2)} = K_A^{(1)}$ and $K_R^{(2)} = K_R^{(1)}\, .$
This assumption implies 
sequential independence of the equilibrium concentrations since  $[BCiRCiA] = [BCiACiR]$.
It is plausible to assume the same relation for later bindings, that is,    $K_A^{(j)} = K_A^{(1)}$ and $K_R^{(j)} = K_R^{(1)}$  for $j\geq 2$ and in 
consequence of this sequential independence  we will denote the dissociation constants by $K_A$ and $K_R$  skipping  the superscript.
Then, if  all the  TFs under consideration are non cooperative we easily deduce that 
the concentration at equilibrium of a configuration with $j_A$ activators and $j_R$ repressors bound is
\begin{equation}\label{gen_eq_conc_bifunc}
[BCiA^{j_A}CiR^{j_R}] = [B] \left(\frac{[CiA]}{ K_A} \right) ^{j_A} \left(\frac{[CiR]}{ K_R} \right) ^{j_R}\,
\end{equation}
independently of the sequential order of binding and  of  the specific positions occupied for the TFs.  
Let us recall that \textit{Drosophila}'s cis-regulatory elements involve a total number of $3$ binding sites, so we have a restriction for the possible number of bound transcription factors. So, $j_A +j_R \leq 3$   has to be verified, and in consequence
$j_0 = 3 - j_A -j_R \geq 0$ denotes the number of free spaces in the configuration.

Cooperation occurs when the existence of other previously bound protein affects to the affinity of the new binding protein of
type  $i,\ h= A,\ R$, that is:
$$K_i^{(2)} = K_h^{(1)}/ c 
$$ 
where $c$ is a positive constant bigger than $1$ if proteins cooperate and smaller than the unity if anti cooperation occurs. Since the only difference between cooperativity and anticooperativity is a threshold value  for $c$ in the subsequent considerations about modelling  we will 
refer to the constant $c$ and not distinguish between both cases.
If cooperation occurs it would be necessary to know which TFs are affected by others TFs since the equilibrium concentration 
will depend on these relations. In the literature total and partial cooperation have recently been proposed to play a very relevant role in the Hh/Shh target 
genes by means of the Ci/Gli TFs \cite{Parker2011,Junker2014}. 
Partial cooperation in the activators would occur when the existence of an activator modify equally the affinity of any posterior activator binding, that is $K_A^{(j)} = K_A^{(1)}/ c_A$  for $j\geq 2$, and respectively for repressors. Total cooperation would occur when the presence of a bound TF modify the affinity of any posterior binding in the same manner, i.e. $K_A^{(j)} = K_A^{(1)}/ c$ and simultaneously $K_R^{(j)} = K_R^{(1)}/ c$  for $j\geq 2$ (see for instance \cite{Lai2004}).
In the subsequent we will denote by $K_A$ or $K_R$ the activator and repressor affinity constants, such that  
\begin{equation}\label{ConcTotalCoop}
[BCiA^{j_A}CiR^{j_R}] = [B] c^{\left(j_A + j_R -1\right)_+} \left(\frac{[CiA]}{ K_A} \right) ^{j_A} \left(\frac{[CiR]}{ K_R} \right) ^{j_R}
\end{equation}
in the presence of total cooperativity while
\begin{equation}\label{ConcPartialCoop}
[BCiA^{j_A}CiR^{j_R}] = [B] c_A^{\left(j_A  -1\right)_+} c_R^{\left(j_R -1\right)_+}  \left(\frac{[CiA]}{ K_A} \right) ^{j_A} \left(\frac{[CiR]}{ K_R} \right) ^{j_R}\, 
\end{equation}
if partial cooperativity for TFs occurs.
Here $(\cdot)_+$ denotes the positive part function ( $(x)_+ =x$ if $x>0$ and zero otherwise) needed because the cooperation will not take place unless two or more 
cooperative TFs are present in the configuration.  In the subsequent we will denote by $\{\{CiA,CiR\}_c\}$ and $\{\{CiA\}_{c_A},\{ CiR\}_{c_R}\}$ the 
total and  partial cooperativity respectively. Let us observe that this notation covers the case of non cooperativity since it would correspond to the 
case $\{\{CiA,CiR\}_1\}$ or equivalently  $\{\{CiA\}_{1},\{CiR\}_{1}\}$.
 
The binding sites are ordered spatially and, in general, there is not an unique spatial distribution  for a configuration with $j_A$ activators, $j_R$ repressors and $3-j_A-j_R$ free sites. For instance if we consider $j_A= j_R= 1$ there are six possible spatial 
distributions with the same elements ($CiACiRO$, $CiRCiAO$, $CiAOCiR$, $CiROCiA$, $OCiACiR$, $OCiRCiA$ where $O$ denotes the empty space).
In our description spatial localization of bound particles is not considered, indeed  for a concrete configuration with $j_A$ activators, $j_R$ repressors and $j_0$ free sites $\frac{3!}{j_0! j_A! j_R!}$ spatial different configurations are plausible, where $k!$ denotes the factorial of $k$.

Regarding the promoter's RNA polymerase binding process, the TFs work together trying to promote or repress the binding process~\cite{Muler2000} by a mechanism  known as recruitment~\cite{Ptashne1997,Ptashne2005}.
Thus, we consider that the activators interact with RNAP with `adhesive' interaction\cite{Bintu2005} that gives rise to a modification of the RNA polymerase binding affinity: $K_{RP}/a^{j_A}$ where where $a$ is a constant bigger than $1$, called for now on activator interaction with the RNA polymerase. On the other hand, the effect of $j_R$ repressors is modelled in terms of a `repulsive' interaction that modifies the binding affinity $K_{RP}/r^{j_R}$ with a factor $r<1$ (repressor interaction). We will refer to these parameters as TFs transcriptional activation/repression intensity.

By using the previous guidelines we are going to describe the concentrations of all possible configurations following \cite{Ackers1982, Shea1985}:

\vspace{0.2cm}
\textbf{Step 1: Construction of the sample space}

Let us consider $[CiA]$, $[CiR]$ and $[RNAP]$ concentrations  of activators, repressors and RNA polymerases. Then,
all the possible ways of obtaining an equilibrium concentration with $j_A$, $j_R$ and $j_P$ activators, repressors and 
RNA polymerases is given by the  states
\begin{eqnarray}\label{ZbifuncRecruitment}
 Z^{(3)}(j_A,j_R,j_P=1;{\mathcal{C}}) 
& = & C(\mathcal{C}) \frac{3!}{j_0! j_A! j_R!} [B] \frac{[RNAP]}{K_{RP}}\left(\frac{a [CiA]}{K_{A}}\right)^{j_A} 
 \left(\frac{r [CiR]}{K_{R}}\right)^{j_R}\,,   \\ 
 Z^{(3)}(j_A,j_R,j_P=0;{\mathcal{C}}) 
& = & C(\mathcal{C}) \frac{3!}{j_0! j_A! j_R!} [B]   \left(\frac{ [CiA]}{K_{A}}\right)^{j_A} 
 \left(\frac{ [CiR]}{K_{R}}\right)^{j_R} \nonumber
\end{eqnarray}
where $j_P=1$ stands for the case where there is a bound RNA polymerase and $j_P=0$ otherwise, 
$j_0 = 3- j_A -j_R \geq 0$, and the variable $\mathcal{C}$ describes the relation of cooperation between the TFs. More concretelly, by using \eqref{ConcTotalCoop} and \eqref{ConcPartialCoop},  the cooperation function $C$ takes the values 
\begin{equation}\label{C_global_cooperationBifunctional}
C(\mathcal{C} = \{ CiA,\ CiR\}_c)=c^{(j_A+j_R-1)_+}
\end{equation}
and
\begin{equation}\label{C_partial_cooperationBifunctional}
C(\mathcal{C} = \{  \{ CiA\}_{c_A},\{ CiR\}_{c_R} \})=c_A^{(j_A-1)_+} c_R^{(j_R-1)_+} \, .
\end{equation}
This allow us to describe all the sample space, i.e. the space of all the possible configurations, as
 $$\Omega= \big\{ (j_A,j_R,j_P)\, ;  \ j_A, j_R \geq 0, \  j_A +j_R \leq 3 , \  j_P = 0,1 \big\}\,.$$

\vspace{0.2cm}
\textbf{Step 2: Definition of the probability}

Once we have described all the possible configurations in terms of the concentrations of activator, repressor and RNA polymerase 
we easily obtain  the probability of finding the promoter in a particular configuration of $j_P$ RNA polymerase and $j_A,\ j_R$ TFs 
related by a cooperation relation $\mathcal{C}$ as
\begin{equation}\label{probability_generalBifunctional}
P^{(3)}(j_A,j_R,j_P; \mathcal{C})=\frac{Z^{(3)}(j_A,j_R,j_P; \mathcal{C})}{\sum\limits_{\{j_A',j_R',j_P'\}\in\Omega}Z^{(3)}(j_A',j_R',j_P';\mathcal{C})}\, ,
\end{equation}
for all  $ (j_A,j_R,j_P) \in \Omega$.

\vspace{0.2cm}
\textbf{Step 3: Definition of the BEWARE operator}

In this last step, the BEWARE operator is obtained in terms of the probabilities $P^{(3)}$. Following the work of Shea et al \cite{Shea1985} the synthesis of certain protein depends on the total probability of finding RNA polymerase in the promoter, more concretely, it is proportional to the marginal distribution of the case $j_P=1$\cite{Bintu2005, Parker2011}. We will denote by the recruitment BEWARE operator to the function
$$
\mbox{BEWARE}([CiA],[CiR],[RNAP];\mathcal{C})
= 
C_B \sum_{j_A,j_R \geq 0}^{j_A+j_R \leq 3} P^{(3)} ( j_A,j_R,j_P =1;\mathcal{C})
$$
where in definition \eqref{probability_generalBifunctional} expression \eqref{ZbifuncRecruitment} is assumed and $C_B$ is a constant of proportionality that could depend on other factors 
disregarded in this work.
Splitting the denominator in two sums, depending on the existence of RNA polymerase bound to the configuration, this expression can be easily rewritten in terms of the regulation factor function, $F_{reg}$:
\begin{eqnarray}
\mbox{BEWARE}([CiA],[CiR],[RNAP];\mathcal{C})
&= & 
\frac{C_B}{1+ \frac{\sum_{j_A',j_R' \geq 0}^{j_A'+j_R' \leq 3} Z^{(3)}(j_A',j_R',j_P'=0;\mathcal{C}) }{\sum_{j_A',j_R' \geq 0}^{j_A'+j_R' \leq 3} Z^{(3)}(j_A',j_R',j_P'=1;\mathcal{C}) }} 
= 
\frac{C_B}{ 1 + \frac{K_{RP}}{[RNAP] F_{reg}([CiA],[CiR];\mathcal{C})}}\, .
\label{beware_recruitmentBifunctional}
\end{eqnarray}

Doing some basic algebra (see Suplemental Material \ref{refining}) this regulation factor can be transformed equivalently into simple expressions whose analysis can contribute to the understanding of the general process. 
These calculations, using a classical strategy employed for obtaining the derivation of the General Binding Equation more than a century ago \cite{Bisswanger2008}, 
have not still been applied in this context up to the authors knowledge.
Indeed, we can prove that the regulation factor can be equivalently written as
\begin{eqnarray}\label{Freg}
F_{reg}([CiA],[CiR]; \mathcal{C}) & = & \frac{S_1^{(3)}\big(a[CiA]K_A^{-1} , r[CiR] K_R^{-1}; \mathcal{C}\big)}{S_1^{(3)}\big([CiA]K_A^{-1}, [CiR] K_R^{-1}; \mathcal{C}\big)}\, ,
\end{eqnarray}
where the explicit expression of $S_1^{(3)}(x,y;\mathcal{C})$ depends on the kind of cooperativity presumed, that is
\begin{eqnarray}
S_1^{(3)}(x,y; \{\{CiA, CiR \}_1\})  & =&   ( 1 + x+y)^3 \, , \label{nc} \\ 
S_1^{(3)}(x,y; \{\{CiA, CiR \}_c\}) & = &  1-\frac{1}{c} + \frac{1}{c}  ( 1 + c x+ c y)^3 \, ,  \label{gc} \\ 
S_1^{(3)}(x,y; \{\{CiA\}_{c_A}, \{CiR \}_{c_R}\})   & = & \frac{(1 + c_A x+c_R y)^3}{c_A c_R}+\left(1-\frac{1}{c_R}\right)  \frac{(1+ c_A x)^3}{c_A} \nonumber  \\ 
&  & + \left(1-\frac{1}{c_A}\right)\frac{(1+ c_R y)^3}{c_R}+\left(1-\frac{1}{c_A}\right)\left(1-\frac{1}{c_R}\right)\, ,  \label{ic}\quad 
\end{eqnarray}
respectively for the non cooperative, total an partial cooperative cases. The first remark is that mathematical complexity in these expressions is mainly related with the assumed cooperativity. 
These ideas can be generalized in a  multifunctional framework and they can be applied in other different contexts as can be seen in~\cite{paperMultifunctional}.

\subsection*{Determination of activation/repression regions}
%
To cope with the problem treated in~\cite{Parker2011,Barolo2013} we are going to establish theoretical regions of activation/repression and posteriorly we will put them in correspondence with biological observations.  Using reporter genes in~\cite{Parker2011,Barolo2013} it was compared the activity of different versions of the \textit{dpp} enhancer containing
three low-affinity sites,  three  high-affinity sites or  three null-affinity sites.
The last one, the basal expression, collects the effects of all other factors than Ci on \textit{dpp} since null-affinity sites interrupt Cubitus control. 
For instance, it takes into account that Engrailed prevents the transcription of \textit{dpp} near the anterior/posterior (A/P) boundary and, in consequence, the basal expression depends on the distance of the cells to this boundary.
More concretely, in\cite{Parker2011,Barolo2013} the effects  of Ci signaling with low- or high-affinity enhancers was measured comparing the gene activity \textit{versus} the basal at any cell.
 We propose  to define activation/repression regions in the plane $[CiA]-[CiR]$ separated by the basal state, that is, we want to determine which concentrations $[CiA],\ [CiR]$  will provide more or less gene expression than the basal.  
Note that the basal state, determined by the absence of TFs, that is $[CiA]=[CiR]=0$, 
 corresponds to $F_{reg}=1$ in expression  (\ref{beware_recruitmentBifunctional}). Thus, the regulation factor could be seen as describing an effective increase (for $F_{reg} >1$) or decrease  (for $F_{reg} <1$) of the number of RNAP molecules bound to the promoter with respect to the basal level as was pointed out in~\cite{Bintu2005}.
The analysis of the influence of the biochemical mechanisms (mainly related with affinities and cooperation) 
on the threshold between both regions will provide us interesting information in order to understand the wide spreading of \textit{dpp}.

Let us first consider  a BEWARE operator with non/total cooperativity ($c\geq 1$) both for \textit{ptc} and for \textit{dpp} obeying the general expression (\ref{beware_recruitmentBifunctional}) which depends 
on the regulation factor
\begin{eqnarray}
F_{reg}([CiA],[CiR];\{CiA,CiR\}_c) 
: = 
\frac{1-\frac1c +\frac1c \left(1+a c\frac{[CiA]}{K_A}+rc\frac{[CiR]}{K_R}\right)^3}{1-\frac1c +\frac1c\left(1+c\frac{[CiA]}{K_A}+c\frac{[CiR]}{K_R}\right)^3}\, . \label{Dpp_beware_nc}
\end{eqnarray}
It is quite easy to see  that in the case of \eqref{Dpp_beware_nc} the threshold 
(that is $F_{reg} =1$)  is determined by the linear relation
\begin{equation}\label{ActRepThreshold}
\frac{[CiR]}{K_R} = \frac{a-1}{1-r}\frac{[CiA]}{K_A}\
\end{equation}
dividing the plane $[CiA]-[CiR]$ into two parts that we can denominate activated region if $\frac{[CiR]}{K_R}<\frac{a-1}{1-r}\frac{[CiA]}{K_A}$ 
 and repressed region if on the contrary $\frac{[CiR]}{K_R}>\frac{a-1}{1-r}\frac{[CiA]}{K_A}$.
Let us observe that the threshold \eqref{ActRepThreshold} is  independent of the affinities values if equal affinities are assumed, while differential affinities between activators and repressors at the common binding sites will modify the slope of this linear threshold. 
See  Fig. \ref{FIG.displaced_threshold} {\bf(a)} and {\bf(c)} where some examples of these thresholds are depicted for different values of the
parameters. 
This trivial remark could give some insight into some experiments where the activation seems to occur at small concentrations of activators even in the presence of a high gradient of repressors (see for instance \cite{Barolo2013}). 

\noindent Moreover, in the absence of cooperativity ($c=1$) the isolines of the regulation factor (contour lines determining values of CiR and CiA concentrations determining the same level of activation)  are the straight lines in the $[CiA]-[CiR]$ plane given by the formula: 
\begin{equation}\label{curvas_de_nivel}
F_{reg}=K\iff \frac{[CiR]}{K_R}=\frac{K^{1/3}-1}{r-K^{1/3}}+\frac{K^{1/3}-a}{r-K^{1/3}}\frac{[CiA]}{K_A}\, .
\end{equation}

In the case of partial cooperativity, however, due to the more entangled expression obtained in (\ref{ic}) it is not so clear that the activation threshold can be obtained explicitly as we did with \eqref{ActRepThreshold} and we need to develop with a little more care the proper mathematical analysis. Indeed, if we impose the threshold equation
$$F_{reg}([CiA],[CiR];\{\{CiA\}_{c_A},\{CiR\}_{c_R}\})=1\, ,$$
it can be shown that this threshold is determined by an unique increasing function $f$, that divides the plane [CiA]-[CiR] in two regions, activation 
($[CiR]/K_R<f([CiA]/K_A)$) and repression ($[CiR]/K_R>f([CiA]/K_A)$)  (see Supplemental Material  \ref{ProofThreshold}  for definition and analysis of the function $f$).
Let us mention that the  threshold of a BEWARE operator with partial cooperativity is not, in general, a straight line although it shows a linear 
asymptotic behaviour for large concentrations (see Fig. \ref{FIG.displaced_threshold} {\bf(e)}).

\section*{Biochemical mechanisms that modify genetic spatial expression}
\begin{figure}[!htb]
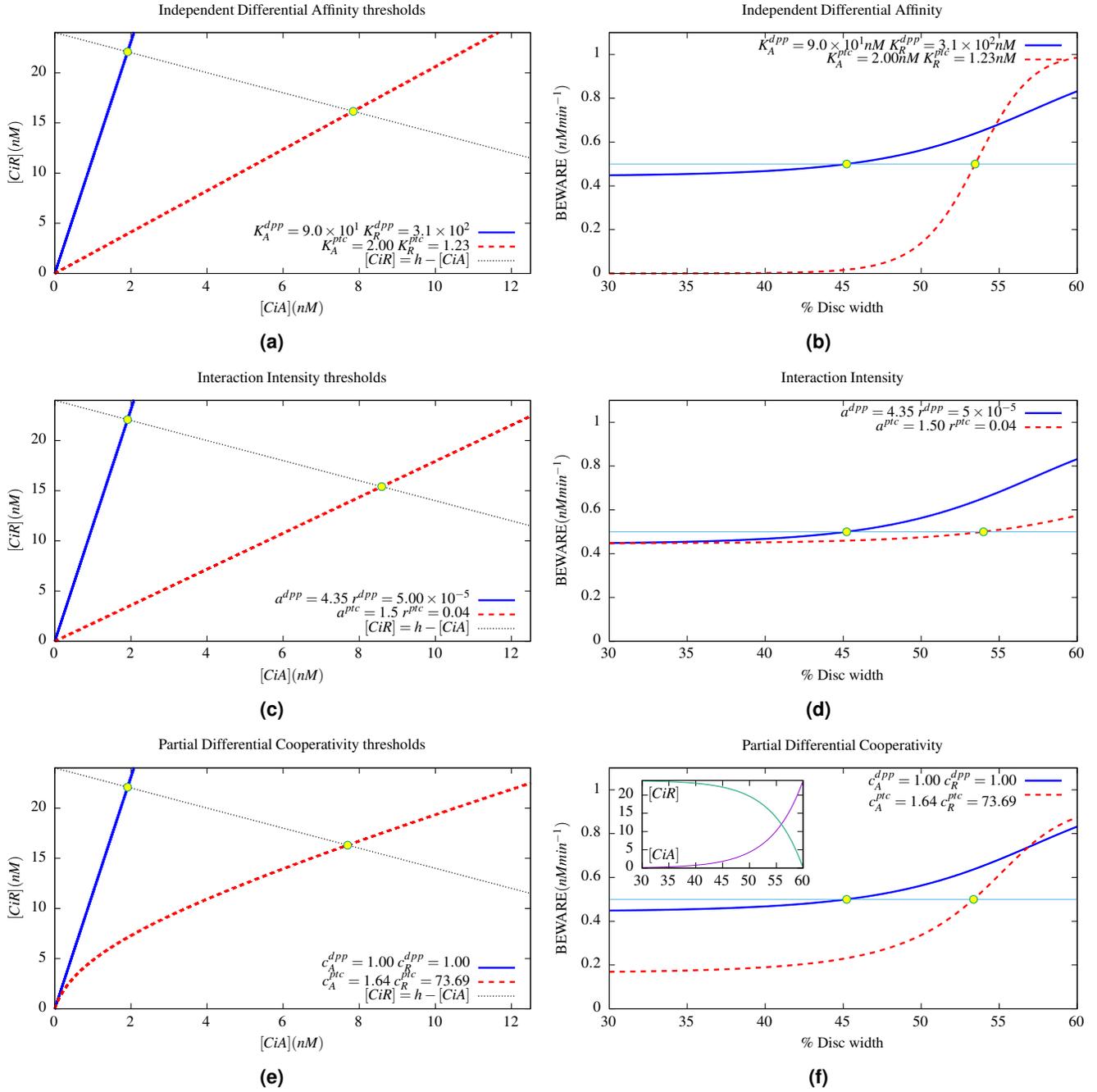

\begin{subfigure}[b]{0.5\textwidth}
\scalebox{0.7}{
\input{thresholds_differential_independent_affinity.tex}
}
\caption{}
\end{subfigure}
\begin{subfigure}[b]{0.5\textwidth}
\scalebox{0.7}{
\input{afinidad_independiente_diferenciada.tex}
}
\caption{}
\end{subfigure}
\begin{subfigure}[b]{0.5\textwidth}
\scalebox{0.7}{
\input{thresholds_differential_control_intensity.tex}
}
\caption{}
\end{subfigure}
\begin{subfigure}[b]{0.5\textwidth}
\scalebox{0.7}{
\input{intensidad_diferenciada.tex}
}
\caption{}
\end{subfigure}
\begin{subfigure}[b]{0.5\textwidth}
\scalebox{0.7}{
\input{thresholds_differential_partial_cooperativity.tex}
}

\caption{}
\end{subfigure}
\begin{subfigure}[b]{0.5\textwidth}
\scalebox{0.7}{
\input{cooperatividad_parcial_diferenciada.tex}
}
\caption{}
\end{subfigure}
\caption{
Biochemical mechanisms changing activation/repression regions and spatial genetic expression.
Blue lines are obtained from a non cooperative BEWARE operator ( \eqref{beware_recruitmentBifunctional}, \eqref{Freg}, \eqref{nc}), 
under {Cubitus} distributions (\eqref{CiADistrib},\eqref{CiTot}) determined by parameters in table \ref{tabla}. Red dashed lines correspond to BEWARE operator involving new biochemical mechanism: Differential Independent Affinity (figures {\bf(a)} and {\bf(b)}), Differential Control Intensity effects ({\bf(c)} and {\bf(d)}) and Differential Partial Cooperativity ({\bf(e)} and {\bf(f)}) with partial cooperative BEWARE operator \eqref{ic}.
Inset in {\bf(f)} shows the opposing gradients of activator and repressor {Ci} considered in this work. 
Yellow circles in  {\bf(a)}, {\bf(c)} and  {\bf(e)} are the intersection points  $([CiA]_{th},[CiR]_{th})$ while in {\bf(b)}, {\bf(d)}, {\bf(f)} are the determined by distances $x_{th}$.
}
\label{FIG.displaced_threshold}
\end{figure}

In this section we will describe the effect induced on the spatial gene expression by the molecular mechanisms: affinity, cooperativity and TFs interaction intensity.  Since our main goal is to understand how these mechanims could modify the expression of the Hh target genes we are going to assume transcription factors acting in the same way that Cubitus works in the \textit{Drosophila} system. We recall that Hh is secreted from the posterior  in the anterior compartment of the wing imaginal disc, that results in opposing gradients of activator and repressor Ci. 
In order to model  these concentration distributions we are going to adopt the time independent approach proposed in\cite{Parker2011}:
\begin{equation}\label{CiADistrib}
 [CiA] = h e^{-x/\sqrt{D}} \, ,
\end{equation} 
where $x$ denotes the distance from the A/P boundary, $h$ scales the activator concentration values and $D$ is the steepness of the gradient.  The A/P boundary is located around the 60\% of the dorso-ventral (D/V) axis and the influence of Hh gradient can be appreciated in the middle of the anterior compartment, more concretely the cells located approximately between the 
30\% and the 60\% of the D/V axis.
 Furthermore, the description used in\cite{Parker2011} also considers the conservation of the total amount of Ci, i.e. 
 \begin{equation} \label{CiTot}
  [CiR]  + [CiA] = h \, , 
 \end{equation}
hence they will be restricted to a  straight line in the $[CiA]-[CiR]$ plane (see Fig. \ref{FIG.displaced_threshold} {\bf(a)}, {\bf(c)}, {\bf(e)}). Inset in Fig. \ref{FIG.displaced_threshold} {\bf(e)} shows the distributions of   $[CiA]\, ,\  [CiR]$ under 
\eqref{CiADistrib} and \eqref{CiTot} that we consider in this work. 
The intersection point between the straight line  \eqref{CiTot}  and the threshold, $([CiA]_{th},[CiR]_{th})$, will determine a 
boundary between genetically activated/repressed cells (represented by yellow circles in Fig. \ref{FIG.displaced_threshold} {\bf(a)}, {\bf(c)}, {\bf(e)}).
That is, repressed (resp. activated) cells will be those containing concentrations $([CiA],[CiR])$ verifying \eqref{CiTot} such that $[CiA] < [CiA]_{th}$ (resp. $[CiA] > [CiA]_{th}$)  and exhibiting, in consequence, transcription rates lower than the basal (resp. higher than the basal). Due to the monotone character of distribution \eqref{CiADistrib} activated cells are closer to the A/P boundary
and the limit of the percentage of the wing imaginal disc occupied by  activated cells will be determined 
by the distance $x_{th}$ verifying
$$[CiA]_{th} = h e^{-x_{th}/\sqrt{D}} \, $$
represented by yellow circles in Fig. \ref{FIG.displaced_threshold} {\bf(b)}, {\bf(d)}, {\bf(f)}.  From now we will refer to the space occupied by activated/repressed cells as relative activated/repressed disc.

By using previous considerations, we will analyse in next paragraphs the effect over the spatial \textit{dpp} and \textit{ptc} expression rate due to the biochemical mechanisms:

\begin{enumerate}
\item Equal Affinity ($K_A^{dpp} = K_R^{dpp}$, $K_A^{ptc} = K_R^{ptc}$).
\item Differential Affinity ($K_A^{dpp} \neq K_R^{dpp}$, $K_A^{ptc} \neq K_R^{ptc}$): In this case we have noticed that it will be relevant to distinguish between Proportional Differential Affinity, where $K_R^{dpp} / K_A^{dpp}= K_R^{ptc}/ K_A^{ptc}$ and Independent Differential Affinity, $K_R^{dpp} / K_A^{dpp} \neq K_R^{ptc}/ K_A^{ptc}$.
\item Interaction intensity, where $a^{dpp} \neq a^{ptc}$, $r^{dpp} \neq r^{Ptc}$.
\item Cooperativity: Global Cooperativity, where $c^{dpp} \neq c^{ptc}$, and Partial Differential Cooperativity, where $c_A^{dpp}\neq c_R^{dpp}$ , $c_A^{ptc}\neq c_R^{ptc}$.
\end{enumerate}
Where the superscripts $ptc$ and $dpp$ stand for the parameters of the genes \textit{ptc} and \textit{dpp}, respectively. Let us remark that some of these mechanisms have been proposed in \cite{Parker2011}.  

Equation (\ref{ActRepThreshold}) shows a clear dependence of the activation-repression threshold (at least for the non-cooperative and total cooperative case) on the activator-repressor affinity constants $K_A$, $K_R$, and activator-repressor interaction intensities $a$ and $r$. 


Under Equal Affinity ( $K_R^{dpp}/K_A^{dpp}= K_R^{ptc}/K_A^{ptc}=1$) the threshold doesn't change. In the case of Differential Affinity the threshold only will change if the affinities are not proportional. Since \textit{dpp} shows a relative activated disc larger than \textit{ptc}, and supposing that this variation comes only from differential affinities, this effect can only be obtained by the non proportional  relation  $K_R^{dpp}/K_A^{dpp}>K_R^{ptc}/K_A^{ptc}$ (see Figures \ref{FIG.displaced_threshold}.a and \ref{FIG.afinidad_global_diferenciada_cooperatividad_global_diferenciada}.a for graphical examples of these effects).

On the other hand, the dependence with $a$ and $r$ in (\ref{ActRepThreshold}) clarifies the main contribution of the activator and repressor interaction intensities to the model: the larger the slope (this is, the larger the value of $a$), the larger will be the relative activation disc, and the larger value of $r$, the less the value of the slope and hence less relative activation disc. In particular, the wide spreading of \textit{dpp}, for the same reason as before, could be motivated by $(a^{dpp}-1)/(1-r^{dpp})>(a^{ptc}-1)/(1-r^{ptc})$ (see Figure \ref{FIG.displaced_threshold}.b for a graphical example of this effect).  
Note also the non-dependence in \eqref{ActRepThreshold} with the cooperation constant $c$, meaning that the activation threshold will remain the same if the TFs cooperate between them in a total manner, no matter the value of the coefficient $c$ is.

\vspace{10pt}
However, if the operator is under the partial cooperation hypothesis \eqref{ic} the expression of the threshold in general does not follow the linear dependence \eqref{ActRepThreshold} because  the partial cooperation constants $c_A$, $c_R$  play an important role in the definition of the activated and repressed states. 
Indeed, in Suplemental material \ref{ProofThreshold} we compare this threshold with the linear one \eqref{ActRepThreshold} obtaining the following result in terms of the magnitudes
\begin{eqnarray}
\bar{a}_ 2 &=& 3 c_A c_R\frac{1-r}{a-1}\big\{ (ar-1) (2-c_A-c_R) +(c_A-c_R) (a-r)  \big\} \label{asubi}\, , \\
\bar{a}_ 3 &=& c_A c_R\frac{1-r}{(a-1)^2}\big\{ (1 +a+a^2) (1-r)^2  c_A^2  + 3 c_A (1-r) (a^2 r-1)  - 3 c_R (a-1) (1-a r^2) - (1 +r+r^2) (a-1)^2  c_R^2 \big\}\, . \nonumber
\end{eqnarray}
Depending on the sign of these values, given by the sign of the terms between brackets, it can be proved that the inclusion of partial cooperativity provokes:
\begin{itemize}
\item If $\bar{a}_2>0$ , $\bar{a}_3>0$: an increment in the activation range with respect to 
the non cooperative case.  
\item If $\bar{a}_2<0$ , $\bar{a}_3<0$: a decrement  of the activation range with respect to 
the non cooperative case.
\item Otherwise: an increment or decrement depending on the total amount of Ci ($h$) considered. A detailed explanation can be found in  Suplemental material \ref{ProofThreshold}. 
\end{itemize} 
See for instance Fig. \ref{FIG.displaced_threshold} {\bf (e)} and {\bf (f)} where the values adopted verify $\bar{a}_2<0$ , $\bar{a}_3<0$ 
and in consequence the activation range has been reduced.

\section*{Biochemical mechanisms that modulate signaling}

Now we are going to prove very easily  the next qualitative property:
 proportional binding affinities  and total cooperation modulate the activation/repression intensity.  However, as we mentioned in the previous section, these biochemical mechanisms do not change the activation range.

Indeed, by using electrophoretic mobility shift assays in vitro, it was measured in ~\cite{Parker2011}  that $Ci^{ptc}$ sites affinities are considerably higher than affinities of $Ci^{dpp}$ sites which in terms of dissociation constants could be interpreted as $K_R^{ptc} << K_R^{dpp}$  and  $K_A^{ptc} << K_A^{dpp}$. Previously, we described how independent differential affinities were able to change the activation regions, and now we are going to see what could be the effect over the transcription levels if proportional differential affinities are assumed, that is, 
\begin{equation} \label{lowhighaffinity}
K_R^{ptc}  = \delta_0K_R^{dpp} \quad \mbox{  and }  K_A^{ptc} =\delta_0K_A^{dpp}\quad  \mbox{ being } \delta_0 << 1\,.
\end{equation}
Since both genes are controlled by the common Ci signaling we can compare the corresponding protein production by comparing the  regulation factors for both genes, that is,
\begin{eqnarray}
F_{reg}^{dpp}
= 
\frac{1-\frac1c +\frac1c \left(1+ac\frac{[CiA]}{K_A^{dpp}}+rc\frac{[CiR]}{K_R^{dpp}}\right)^3}{1-\frac1c +\frac1c\left(1+c\frac{[CiA]}{K_A^{dpp}}+c\frac{[CiR]}{K_R^{dpp}}\right)^3} \nonumber
\end{eqnarray}
versus
\begin{eqnarray}
F_{reg}^{ptc}
=
\frac{1-\frac1c +\frac1c \left(1+ac\frac{[CiA]}{K_A^{ptc}}+rc\frac{[CiR]}{K_R^{ptc}}\right)^3}{1-\frac1c +\frac1c\left(1+c\frac{[CiA]}{K_A^{ptc}}+c\frac{[CiR]}{K_R^{ptc}}\right)^3}
= 
\frac{1-\frac1c +\frac1c \left(1+ac\frac{[CiA]}{\delta_0 K_A^{dpp}}+rc\frac{[CiR]}{\delta_0 K_R^{dpp} }\right)^3}{1-\frac1c +\frac1c\left(1+c\frac{[CiA]}{ \delta_0 K_A^{dpp}}+c\frac{[CiR]}{\delta_0 K_R^{dpp}}\right)^3}\, 
 \nonumber
\end{eqnarray}
where relation \eqref{lowhighaffinity} has been replaced. Let us remark that this can be done thanks to the fact that the BEWARE operator \eqref{beware_recruitmentBifunctional} is monotone with respect to the Regulation Factor.
It can be shown, by some monotonicity properties of these operators (see Supplemental Material \ref{monotonous_proof}), that
$$F_{reg}^{ptc}>F_{reg}^{dpp}(>1)\, ,\mbox{ if }[CiA]\mbox{ and }[CiR]\mbox{ belong to the activation region}$$
\vspace{0.2cm}
and 
\vspace{0.2cm}
$$F_{reg}^{ptc}<F_{reg}^{dpp}(<1)\, ,\mbox{ if } [CiA]\mbox{ and }[CiR]\mbox{ belong to the repression region}.$$
%
This result can be interpreted in the following terms:
binding affinity reduction (increment in $\delta$) provokes less activation in the activation region and less repression in the repression region. That is, for low-affinity binding sites the signaling is attenuated. This effect can be clearly observed in previous literature, more concretely in \cite{Parker2011} Figure S6 (C)  where a fit of a biophysical model of BEWARE recruitment type considering non cooperativity ($c=1$)  and equal affinity ($K_R= K_A$)  were fitted to high-affinity ($3\times Ci^{ptc}$) and  low-affinity ($3\times Ci^{wt}$) data.  In this figure the threshold between activation/repression coincide in both fittings and the relative expression of the low-affinity fitting is attenuated with respect to the high-affinity one.

\begin{figure}[!htb]
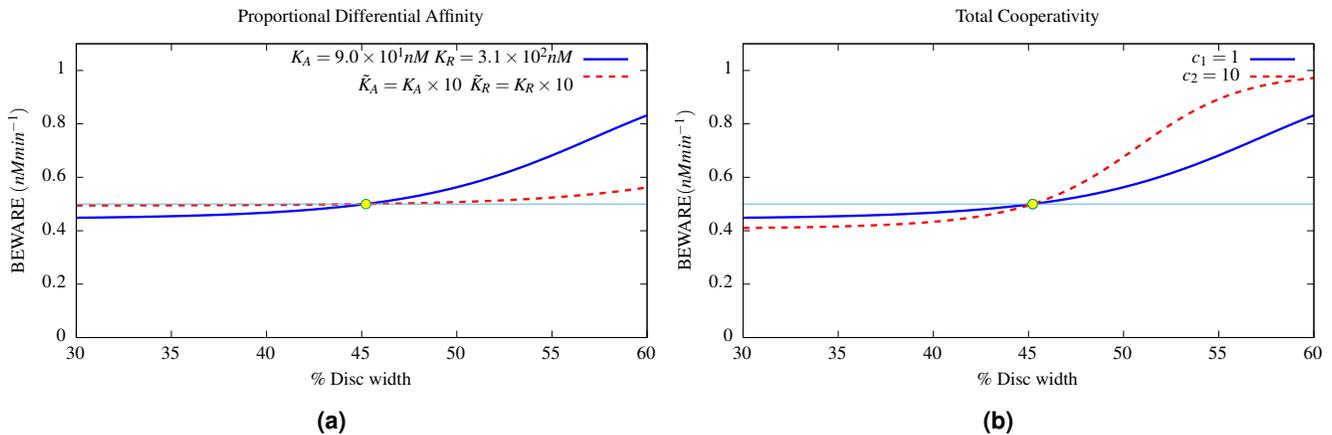

\begin{subfigure}[b]{0.5\textwidth}
\scalebox{0.7}{
\input{afinidad_global_diferenciada.tex}
}
\caption{}
\end{subfigure}
\begin{subfigure}[b]{0.5\textwidth}
\scalebox{0.7}{
\input{cooperatividad_global_diferenciada.tex}
}
\caption{}
\end{subfigure}
\caption{Biochemical mechanisms that only attenuate or reinforce signaling. Blue lines are obtained from a non cooperative BEWARE operator ( \eqref{beware_recruitmentBifunctional}, \eqref{Freg}, \eqref{nc}), 
under Cubitus distributions (\eqref{CiADistrib},\eqref{CiTot}) determined by parameters in table \ref{tabla}. Red dashed lines correspond to BEWARE operator involving the biochemical mechanism (tilda parameters): Differential Total Affinity ({\bf(a)}) and Differential Total Cooperativity effects ({\bf(b)}) with total cooperative BEWARE operator \eqref{gc}.}
\label{FIG.afinidad_global_diferenciada_cooperatividad_global_diferenciada}
\end{figure}

On the other hand, in the previous section we discussed that the total cooperation does not modify the threshold either. In fact we can prove (see Supplemental Material \ref{monotonous_proof}) that cooperativity strengthens the signaling in the sense that, for $1 \leq c_1 \leq  c_2$,
$$
F_{reg}([CiA],[CiR];\{CiA,CiR\}_{c_1}) \leq F_{reg}([CiA],[CiR];\{CiA,CiR\}_{c_2})
$$
when $[CiA], \ [CiR]$ lay in the activation region while 
$$
F_{reg}([CiA],[CiR];\{CiA,CiR\}_{c_1}) \geq F_{reg}([CiA],[CiR];\{CiA,CiR\}_{c_2})
$$
in the opposite case. That is, the increment on the value of total cooperativity provokes more activation/repression in their corresponding regions.  See Figure \ref{FIG.afinidad_global_diferenciada_cooperatividad_global_diferenciada} for two examples of both attenuation (Fig \ref{FIG.afinidad_global_diferenciada_cooperatividad_global_diferenciada} (\textbf{a}) and reinforcement (Fig \ref{FIG.afinidad_global_diferenciada_cooperatividad_global_diferenciada} (\textbf{b}) effects.

\section*{Discussion and conclusion}

The modelling proposed in this paper provides mathematically treatable BEWARE modules enclosing two well accepted
theoretical approaches in the gene transcription modelling framework: the statistical thermodynamic method and the recruitment mechanism. 
These expressions are susceptible of being tested in deep detail by using mathematical analysis. So they are useful tools for unravelling the complex balances that control the transcription process in 
gene expression. 
This is a methodology that has been fruitfully applied in many other aspects of quantitative biology. 

The particular application of these ideas to the problem of differential spatial activation of the Hh target genes 
predicts that:
\begin{itemize}
\item some a priory hypotheses (proportional low-high affinity and total cooperation) should be disregarded as unique responsible of the change of spatial expression.   
\item other a priory hypotheses (differential affinity and partial cooperation) are available alternatives to be tested,
\item new hypothesis (interaction intensity of the TFs ) could be taken into consideration, although the biological  interpretations of 
this point should be evaluated.
\end{itemize}
The contrast with biological evidences, probably from a different point of view, will be necessary to improve the 
theoretical understanding of this particular problem. From a more broader point of view, the repercussion of the analysis of these functionals may be of deeper bearing as soon as this would be performed in a model focus on the mechanisms controlling the balance between the transcription factors\cite{Casares2013,Dassow2000,Saha2006,Oscar2013}.

\section*{Acknowledgements}
O. S. would like to thanks professor J. Garc\'ia-Ojalvo for pointing out reference \cite{Bintu2005}. This work has been partially supported by the MINECO-Feder (Spain) research grant number MTM2014-53406-R, the Junta de Andaluc\'ia (Spain) Project FQM 954, and the MINECO (Spain) research grant FPI2015/074837 (M.C.).

\newpage

\section{SUPPLEMENTAL MATERIAL}

\subsection{Refining the regulation factor expressions}\label{refining}

In this subsection we present the different procedures used in order to rewrite the regulation factor used in the BEWARE operator (\ref{beware_recruitmentBifunctional}).We will work with two  transcription factors: $[CiA]$ and $[CiR]$, with activation and repression interaction intensity constants $a$ and $r$, and binding affinities $K_A$ and $K_R$. The number of enhancers in the promoter will be denoted by $n$.

\subsubsection{Basic tools}

%
%
%
%

In the next subsection we will show a procedure for obtaining the sums:
\begin{equation}\label{Cooperative_species_sum 2 species}
S_{x_0}^{(n)}(x_A,x_R;\mathcal{C}) = \sum\limits_{j_A,j_R\geq 0}^{j_A+j_R\leq n}C(\mathcal{C}) \frac{n!}{j_0! j_A! j_R!} x_0^{j_0}x_A^{j_A} x_R^{j_R} \quad \mbox{ being } \quad j_0=n-j_A-j_R
\end{equation}
appearing in regulation factor definition \eqref{Freg} by using the Multinomial theorem: 
\begin{equation}\label{multinomial_theorem 2 species}
\left(x_0+x_A+x_R\right)^n=\sum\limits_{j_0+j_A+j_R=n}\frac{n!}{j_0!j_A!j_R!}x_0^{j_0}x_A^{j_A}x_R^{j_R}
\quad \mbox{ where } \quad   j_0\,,  j_A\,,  j_R\ge0\, .
\end{equation}


Indeed, recalling expression (\ref{beware_recruitmentBifunctional}), the BEWARE operator is written in terms of the Regulator  Factor
$$F_{reg}([CiA],[CiR];\mathcal{C})=\frac{\sum\limits_{j_A,j_R\geq 0}^{j_A+j_R\leq n}C(\mathcal{C}) \frac{n!}{j_0! j_A! j_R!} \left(\frac{a [CiA]}{K_{A}}\right)^{j_A} \left(\frac{r [CiR]}{K_{R}}\right)^{j_R}}{\sum\limits_{j_A,j_R\geq 0}^{j_A+j_R\leq n}C(\mathcal{C}) \frac{n!}{j_0! j_A! j_R!} \left(\frac{[CiA]}{K_{A}}\right)^{j_A} \left(\frac{[CiR]}{K_{R}}\right)^{j_R}}=\frac{S_{1}^{(n)}(aK_A^{-1}[CiA],rK_R^{-1}[CiR];\mathcal{C})}{S_{1}^{(n)}(K_A^{-1}[CiA],K_R^{-1}[CiR];\mathcal{C})}\, .$$
We will get  several expression for the  regulation factor  depending on the cooperation between TFs, i.e., non-cooperative \eqref{nc}, total-cooperative \eqref{gc} and partial-cooperative \eqref{ic},  represented by $C(\mathcal{C})$.
Next lines show the computations for the numerators of the regulation factors, for each cooperation hypothesis, and the denominators will follow the same deduction imposing $a=r=1$.

\begin{itemize}
\item \textbf{Regulation factor for non-cooperative species}

\vspace{0.5cm}

In this case the $C(\mathcal{C})=1$ and by (\ref{multinomial_theorem 2 species}) the sum reads 

\begin{eqnarray*}
\lefteqn{
S_{1}^{(n)}(aK_A^{-1}[CiA],rK_R^{-1}[CiR];\{CiA,CiR\}_1)} \\
& = & \sum\limits_{j_A,j_R\geq 0}^{j_A+j_R\leq n}\frac{n!}{j_0! j_A! j_R!} \left(\frac{a [CiA]}{K_{A}}\right)^{j_A} \left(\frac{r [CiR]}{K_{R}}\right)^{j_R} = \sum\limits_{j_0+j_A+j_R=n}\frac{n!}{j_0! j_A! j_R!} 1^{j_0}\left(\frac{a [CiA]}{K_{A}}\right)^{j_A} \left(\frac{r [CiR]}{K_{R}}\right)^{j_R} \\
& = & (1+aK_A^{-1}[CiA]+rK_R^{-1}[CiR])^n \nonumber\text{ .}
\end{eqnarray*}


\item \textbf{Regulation factor for total-cooperative species}

If the transcription factors cooperate between all of them, then the cooperation function is described by (\ref{C_global_cooperationBifunctional}). The cooperation function makes a bit more difficult the calculus of the polynomial (\ref{multinomial_theorem 2 species}), and first we need to get rid of the cooperation in both the numerator and denominator of the regulation factor. This can be easily achieved by splitting the sum

\begin{eqnarray*}
\lefteqn{
S_{1}^{(n)}(aK_A^{-1}[CiA],rK_R^{-1}[CiR];\{CiA,CiR\}_c) 
= 
\sum\limits_{j_A,j_R\geq 0 }^{j_A+j_R\leq n}\frac{n!}{j_0! j_A! j_R!} c^{(j_A +j_R-1)_+} \left(\frac{a [CiA]}{K_{A}}\right)^{j_A} \left(\frac{r [CiR]}{K_{R}}\right)^{j_R}}
\\& 
= 
& 
 1+  \sum\limits_{j_A,j_R\geq 1 }^{j_A+j_R\leq n}\frac{n!}{j_0! j_A! j_R!} c^{(j_A +j_R-1)_+} \left(\frac{a [CiA]}{K_{A}}\right)^{j_A} \left(\frac{r [CiR]}{K_{R}}\right)^{j_R} 
\\ 
& = & 1+ \frac{1}{c}\sum\limits_{j_A,j_R\geq 1}^{j_A+j_R\leq n}\frac{n!}{j_0! j_A! j_R!} \left(\frac{ca [CiA]}{K_{A}}\right)^{j_A} \left(\frac{cr [CiR]}{K_{R}}\right)^{j_R}  
\\&
 = 
 & 
 1-\frac{1}{c} + \frac{1}{c}\sum\limits_{j_A,j_R\geq 0}^{j_A+j_R\leq n}\frac{n!}{j_0! j_A! j_R!} \left(\frac{ca [CiA]}{K_{A}}\right)^{j_A} \left(\frac{cr [CiR]}{K_{R}}\right)^{j_R}\nonumber\text{ ,}
\end{eqnarray*}
and using (\ref{multinomial_theorem 2 species}) as before

$$S_{1}^{(n)}(aK_A^{-1}[CiA],rK_R^{-1}[CiR];\{CiA,CiR\}_c)=1-\frac{1}{c}+\frac{1}{c}(1+caK_A^{-1}[CiA]+crK_R^{-1}[CiR])^n\, .$$


\item \textbf{Regulation factor for partial-cooperative species}

If the TFs cooperate independently (eq. (\ref{C_partial_cooperationBifunctional})), we can split the sum twice in the same way as in the total-cooperation case, i.e.,

\begin{eqnarray*}
\lefteqn{S_{1}^{(n)}(aK_A^{-1}[CiA],rK_R^{-1}[CiR];\{\{CiA\}_{c_A},\{CiR\}_{c_R}\}) = \sum\limits_{j_A,j_R\geq 0 }^{j_A+j_R\leq n}\frac{n!}{j_0! j_A! j_R!} c_A^{(j_A-1)_+} c_R^{(j_R-1)_+} \left(\frac{a [CiA]}{K_{A}}\right)^{j_A} \left(\frac{r [CiR]}{K_{R}}\right)^{j_R}}\\  
& = & \sum\limits_{j_A, j_R\geq 1}^{j_A+j_R\leq n}\frac{n!}{j_0! j_A! j_R!} c_A^{(j_A-1)_+} c_R^{(j_R-1)_+} \left(\frac{a [CiA]}{K_{A}}\right)^{j_A} \left(\frac{r [CiR]}{K_{R}}\right)^{j_R} + \sum\limits_{\substack{j_A \equiv 0 \\ j_R\geq 1}}^{j_R\leq n}\frac{n!}{j_0! j_R!} c_R^{(j_R-1)_+} \left(\frac{r [CiR]}{K_{R}}\right)^{j_R} \\
& + & \sum\limits_{\substack{j_A\geq 1 \\ j_R \equiv 0}}^{j_A\leq n}\frac{n!}{j_0! j_A!} c_A^{(j_A-1)_+}  \left(\frac{a [CiA]}{K_{A}}\right)^{j_A} + 1 \\
& = & \frac{1}{c_A c_R}\sum\limits_{j_A , j_R\geq 1}^{j_A+j_R\leq n}\frac{n!}{j_0! j_A! j_R!} \left(\frac{c_A a [CiA]}{K_{A}}\right)^{j_A} \left(\frac{c_R r [CiR]}{K_{R}}\right)^{j_R} + \frac{1}{c_R}\sum\limits_{\substack{j_A \equiv 0 \\ j_R\geq 1}}^{j_R\leq n}\frac{n!}{j_0! j_R!} \left(\frac{c_R r [CiR]}{K_{R}}\right)^{j_R} \\
& + & \frac{1}{c_A}\sum\limits_{\substack{j_A\geq 1 \\ j_R \equiv 0}}^{j_A\leq n}\frac{n!}{j_0! j_A!} \left(\frac{c_A a [CiA]}{K_{A}}\right)^{j_A} + 1 \\
& = & \frac{1}{c_A c_R}\sum\limits_{j_A , j_R\geq 0}^{j_A+j_R\leq n}\frac{n!}{j_0! j_A! j_R!} \left(\frac{c_A a [CiA]}{K_{A}}\right)^{j_A} \left(\frac{c_R r [CiR]}{K_{R}}\right)^{j_R} + \left(1-\frac{1}{c_A}\right)\frac{1}{c_R}\sum\limits_{\substack{j_A \equiv 0 \\ j_R\geq 1}}^{j_R\leq n}\frac{n!}{j_0! j_R!} \left(\frac{c_R r [CiR]}{K_{R}}\right)^{j_R} \\
& + & \left(1-\frac{1}{c_R}\right)\frac{1}{c_A}\sum\limits_{\substack{j_A\geq 1 \\ j_R \equiv 0}}^{j_A\leq n}\frac{n!}{j_0! j_A!} \left(\frac{c_A a [CiA]}{K_{A}}\right)^{j_A} + 1-\frac{1}{c_A c_R}
\end{eqnarray*}

where, by the previous deduction,

\begin{eqnarray}\label{sumatorio_dos_especies_ic}
\lefteqn{
S_{1}^{(n)}(aK_A^{-1}[CiA],rK_R^{-1}[CiR];\{\{CiA\}_{c_A},\{CiR\}_{c_R}\})} \\
& = & \frac{1}{c_Ac_R}(1+c_AaK_A^{-1}[CiA]+c_RrK_R^{-1}[CiR])^n+\left(1-\frac{1}{c_R}\right) \frac{(1+c_AaK_A^{-1}[CiA])^n}{c_A} \nonumber \\ 
& + & \left(1-\frac{1}{c_A}\right)\frac{(1+c_RrK_R^{-1}[CiR])^n}{c_R}+\left(1-\frac{1}{c_A}\right)\left(1-\frac{1}{c_R}\right) \, .\nonumber 
\end{eqnarray}


\end{itemize}

\subsection{Analysis of the threshold for the BEWARE operator with partial cooperativity}
\label{ProofThreshold}

In this subsection, as we announced at the end of  Section Methods, we develop the mathematical analysis of the function $f$ defining the threshold determined by 
$$F_{reg}([CiA],[CiR];\{\{CiA\}_{c_A},\{CiR\}_{c_R}\})=1\, .$$
More concretly we are going to:
\begin{itemize}
\item prove that the threshold is a regular increasing curve in the plane 
$[CiA]-[CiR]$ by using the implicit function theorem. 
\item describe the behaviour of this threshold.
\end{itemize}
Imposing the previous threshold condition we end up with the equivalent polynomial equation of the form

\begin{equation}\label{partial_cooperative_threshold_implicit}
G\left(\frac{[CiA]}{K_A},\frac{[CiR]}{K_{R}}\right)=0
\end{equation}
being
\begin{eqnarray*}
G\left(\tilde{A},\tilde{R}\right) & = & \left(1+c_A a\tilde{A}+c_R r\tilde{R}\right)^3-\left(1+c_A \tilde{A}+c_R \tilde{R}\right)^3 
\\
& &  
+ (c_R-1)\left[\left(1+ c_A a \tilde{A}\right)^3-\left(1+ c_A \tilde{A}\right)^3\right] 
 +   (c_A-1)\left[\left(1+ c_R r \tilde{R}\right)^3-\left(1+ c_R \tilde{R}\right)^3\right]\, .
\end{eqnarray*}
Please note that, by definition,  $G$ takes negative values in the repression region and  positive values in the activation region. 

We are going to prove that $G$ fulfils the hypothesis of the implicit function theorem. With some basic calculations we rewrite the function $G$ as a polynomial in the $\tilde{R}=\frac{[CiR]}{K_R}$ repression variable, and $\tilde{A}=\frac{[CiA]}{K_{A}}$ dependent coefficients,
\begin{equation}\label{polinomio_threshold}
G(\tilde{A},\tilde{R})\equiv P(\tilde{R}) = a_0(\tilde{A})+a_1(\tilde{A})\tilde{R}+a_2(\tilde{A})\tilde{R}^2+a_3(\tilde{A})\tilde{R}^3
\end{equation}
with 
\begin{equation*}
\left\{
\begin{array}{lll}
a_0(\tilde{A}) & = & c_R\left[(1+ac_A \tilde{A})^3-(1+c_A \tilde{A})^3\right] \vspace{0.2cm} \\ 
a_i(\tilde{A}) & = & \frac{3!}{(3-i)!i!}{c_R}^i\left[r^i(1+ac_A \tilde{A})^{3-i}-(1+c_A\tilde{A})^{3-i}+(c_A-1)(r^i-1)\right] \,\, \forall i=1,2,3 \, .\\
\end{array}
\right.
\end{equation*}
Here we state some lemmas allowing us to employ the implicit function theorem.
\begin{lemma}
Let $a>1$, $r<1$ and $c_A, c_R \geq 1$. Then, for any positive value  $\tilde{A}$, $P(\tilde{R})$ has an unique positive root, $\tilde{R}^*$,  and  $P'(\tilde{R}^*) = \frac{\partial G}{\partial \tilde{R}} (\tilde{A},\tilde{R}^*)< 0$. \\
\end{lemma}

{\bf Proof.} First, note that $a_0>0$ and $a_3=c_R^3 c_A(r^3-1)<0$, due to the hypothesis on the parameters $a>1$ and $r<1$, and $\tilde{A}>0$. Then, it is clear that

$$\lim\limits_{\tilde{R}\rightarrow 0}P(\tilde{R})=a_0>0\, \quad\mbox{ and }\quad \lim\limits_{\tilde{R}\rightarrow \infty}P(\tilde{R})=-\infty\, ,$$
and hence there exist at least one positive root of $P(\tilde{R})$. Note also that, if $P$ has no real extrema, then the result is trivially verified. If there exist real extrema of $P$, then their sign will provide information about the number of roots.  In the cases of pairs of positive-negative and negative-negative extrema, it can be easily checked the existence of a unique positive root, verifying the result. In the remaining case, the existence of two positive extrema would imply
 the existence of three positive roots. We are going to prove that this case cannot be achieved with the conditions of the parameters that the polynomial works with. 



The hypothesis of two real possitive extrema would imply
\begin{equation} \label{condicionraicespositivas}
\left\{
\begin{array}{l}
\tilde{R}_0^{(+)}>0\iff -a_2+\sqrt{a_2^2-3a_3 a_1}<0\, , \\
\tilde{R}_0^{(-)}>0\iff -a_2-\sqrt{a_2^2-3a_3 a_1}<0\, , \\
\end{array}
\right.
\quad \mbox{being} \quad
\tilde{R}_0^\pm=\frac{-a_2 \pm \sqrt{a_2^2-3a_3 a_1}}{3 a_3}\, ,
\end{equation}
due to $a_3<0$ and assuming $a_2^2-3a_3 a_1 \geq 0$.
For the same reason $\tilde{R}_0^{(-)} \geq \tilde{R}_0^{(+)}$ and in consequence  condition \eqref{condicionraicespositivas} can be equivalently written as
\begin{equation}\label{condicion_R0}
\sqrt{a_2^2-3a_3 a_1}< a_2\, .
\end{equation}
From \eqref{condicion_R0} we easily deduce that
\begin{equation}\label{condicion_a2>0}
a_2>0
\end{equation}
and 
\begin{equation}\label{condicion_a1<0}
a_1< 0
\end{equation}
are  necessary conditions  because $a_3 < 0$. 
Let us prove that both conditions, \eqref{condicion_a2>0} and \eqref{condicion_a1<0},  are not compatible and in consequence
\eqref{condicionraicespositivas} can not be verified.

\eqref{condicion_a2>0} can be written equivalently as $ (r^2 a-1)\tilde{A}+r^2-1>0$ , which holds if and only if
\begin{equation}\label{condicion_a2>0_separada_1}
r^2 a>1
\quad \mbox{and} \quad
\tilde{A}>\frac{1-r^2}{r^2 a-1}
\end{equation}
are simultaneously fulfilled.





On the other hand, condition \eqref{condicion_a1<0} can be explicitly expressed as
\begin{equation*}
(ra^2-1)c_A\tilde{A}^2+2(ra-1)\tilde{A}+r-1<0\, .
\end{equation*}
In particular,  when $r^2 a>1$,  this inequality requieres
$2(ra-1)\tilde{A}+r-1<0$ 
which occurs if and only if 
\begin{equation}\label{neces_a1<0}
\tilde{A}<\frac{1-r}{2(ra-1)}.
\end{equation}
Now, let us observe that necessary conditions \eqref{condicion_a2>0_separada_1} and \eqref{neces_a1<0}, respectively for \eqref{condicion_a2>0} and \eqref{condicion_a1<0}, are not compatible at all since $r<1$, $r^2 a>1$ and thus:
$$
\frac{1-r^2}{r^2 a-1} > \frac{1-r^2}{2(r^2 a-1)} >  \frac{1-r}{2(r a-1)} \, .
$$
\hfill $\Box$

In an absolutely symmetric manner we can prove the analogous result fixing the variable $\tilde{R}$ and the corresponding polynomial $\bar{P}(\tilde{A}) = G(\tilde{A},\tilde{R})$. 
\begin{lemma}
Let $a>1$, $r<1$ and $c_A, c_R \geq 1$. Then, for any  positive value $\tilde{R}$, $\bar{P}(\tilde{A})$ has an unique positive root, $\tilde{A}^*$,  and  $\bar{P}'(\tilde{A}^*) = \frac{\partial G}{\partial \tilde{A}} (\tilde{A}^*,\tilde{R})> 0$. \\
\end{lemma}

Both results allow us to define a bijective function such that $f(\tilde{A}) = \tilde{R}^*$ and $f^{-1}(\tilde{R}) = \tilde{A}^*$ because of the 
uniqueness of the roots of $P(\tilde{R})$ and $\bar{P}(\tilde{A})$.  The implicit function theorem gives that $f$ is regular and increasing, since,  $G(\tilde{A},f(\tilde{A}))=0$ for all   $\tilde{A} >0$ then
$$
0 = \frac{\partial G}{\partial \tilde{A} }\big(\tilde{A},f(\tilde{A})\big) +
\frac{\partial G}{\partial \tilde{R} }\big(\tilde{A},f(\tilde{A})\big) f'(\tilde{A}) = 
\bar{P}' (\tilde{A}) + P' (f(\tilde{A})) f'(\tilde{A})
\quad \Longrightarrow  \quad
f'(\tilde{A}) = -\frac{\bar{P}' (\tilde{A})}{ P' (f(\tilde{A}))} > 0\, .
$$
Indeed, the threshold could be computed explicitly by applying the classical Tartaglia-Cardano's method (see \cite{Abramovitz1972}  Secc. 3.8.). However, we can show without using these explicit expressions that $f$ tends asymptotically to a straight line as the concentration of the TFs increases. This can be easily shown by simply evaluating the threshold condition \eqref{partial_cooperative_threshold_implicit} on $\tilde{R}=f(\tilde{A})$ and dividing by $\tilde{A}^3$ the whole equation. Then, tending the activators concentration to infinity leads to the equation

\begin{eqnarray*}
&\lim\limits_{\tilde{A}\rightarrow \infty}& \left(c_A a+c_R r \frac{f(\tilde{A})}{\tilde{A}} \right)^3-\left(c_A+c_R \frac{f(\tilde{A})}{\tilde{A}}\right)^3+ (c_R-1)\left[\left(c_A a\right)^3-c_A^3\right]  +   (c_A-1)\left[\left(c_R r \frac{f(\tilde{A})}{\tilde{A}}\right)^3-\left(c_R \frac{f(\tilde{A})}{\tilde{A}}\right)^3\right] \\
& = & \left(c_A a+c_R r \alpha \right)^3-\left(c_A+c_R \alpha\right)^3+ (c_R-1)\left[\left(c_A a\right)^3-c_A^3\right]  +   (c_A-1)\left[\left(c_R r \alpha\right)^3-\left(c_R \alpha\right)^3\right]=0\, ,
\end{eqnarray*}
where $\alpha = \lim\limits_{\tilde{A}\rightarrow\infty} \frac{f(\tilde{A})}{\tilde{A}}$ is the slope that needs to be finite in the limit in order to fullfill the equation, and hence in the limit $f'(\tilde{A})\rightarrow \alpha$. 

Indeed, one can compare this threshold with \eqref{ActRepThreshold} by simply evaluating $G$ in the straight line \eqref{ActRepThreshold} since

\begin{equation*}
G\left(\tilde{A},\frac{1-a}{r-1}\tilde{A}\right)=\bar{P}(\tilde{A})=(\bar{a}_2+\bar{a}_3\tilde{A})\tilde{A}^2\, ,
\end{equation*}
where the coefficients  $\bar{a}_ 2$ and $\bar{a}_ 3$ are given by \eqref{asubi}.
This convoluted relation between the parameters defines the activation-repression range compared to the linear threshold, where depending on the sign of the coefficients we get:
\begin{itemize}
 \item If $\bar{a}_2 > 0$ and $\bar{a}_3 > 0$, then the threshold for the BEWARE operator with partial cooperativity is over the threshold \eqref{ActRepThreshold}. In consequence, the activation range will increase due to partial cooperativity.
 \item If $\bar{a}_2 < 0$ and $\bar{a}_3 < 0$, then the threshold for the BEWARE operator with partial cooperativity is under the threshold   \eqref{ActRepThreshold}. In consequence, the activation range will decrease due to partial cooperativity.
\item If $\bar{a}_2 < 0$ and $\bar{a}_3 > 0$, the threshold for the  the BEWARE operator with partial cooperativity is over the threshold \eqref{ActRepThreshold}  if  $\tilde{A} = \frac{[CiA]}{K_A} < \frac{-\bar{a}_2 }{\bar{a}_3 }$  and is under \eqref{ActRepThreshold}   otherwise.  In this case, the change in the activation range will be determined by the total level of  Ci protein in the system, $h$, by 
eq.\eqref{CiTot}. 
Let us consider  $$[CiA]_{th}^l = \frac{h}{\frac{a-1}{1-r} \frac{K_R}{K_A} +1}$$  the intersection point between \eqref{ActRepThreshold} and \eqref{CiTot} and $[CiA]_{th}$  the intersection point between the threshold for the  BEWARE operator with partial cooperativity. 

Then, if $[CiA]_{th}^l < \frac{-\bar{a}_2 }{\bar{a}_3 }$ it can be easily checked that $[CiA]_{th} < [CiA]_{th}^l$. On the other hand, when, $[CiA]_{th}^l > \frac{-\bar{a}_2 }{\bar{a}_3 }$  the reverse inequality holds  $[CiA]_{th} > [CiA]_{th}^l$. That is, 
the activation range is larger or shorter with partial cooperativity depending on  $h$. 
\item If $\bar{a}_2 > 0$ and $\bar{a}_3< 0$, the situation is exactly opposite to the previous one. Now, if 
$[CiA]_{th}^l > \frac{-\bar{a}_2 }{\bar{a}_3 }$ then $[CiA]_{th}$ verifies $\frac{-\bar{a}_2 }{\bar{a}_3 } < [CiA]_{th} < [CiA]_{th}^l$
and the activation range will increase. Furthermore, if $[CiA]_{th}^l < \frac{-\bar{a}_2 }{\bar{a}_3 }$ then the activation range will decrease because $\frac{-\bar{a}_2 }{\bar{a}_3 } > [CiA]_{th} > [CiA]_{th}^l$ holds.
 \end{itemize}

%
%

\subsection{Monotonous behaviours for the total cooperation case}\label{monotonous_proof}

In the main work we have stated that the regulation factor is monotonous decreasing and increasing with respect the affinity constant in the activated and repressed zones, i.e.,

\begin{lemma}
Let us consider the function
$$
G (\delta)  = 
\frac{\left(1-\frac1c\right)  +\frac1c \left(1+a\frac{c[CiA]}{\delta K_A}+r\frac{c[CiR]}{\delta K_R }\right)^n}{\left(1-\frac1c\right)  +\frac1c\left(1+c\frac{[CiA]}{\delta K_A}+c\frac{[CiR]}{\delta K_R}\right)^n}
= 
\frac{\left(1-\frac1c\right)\delta^n  +\frac1c \left(\delta+ac\frac{[CiA]}{K_A}+rc\frac{[CiR]}{K_R }\right)^n}{\left(1-\frac1c\right) \delta^n  +\frac1c\left(\delta+c\frac{[CiA]}{K_A}+c\frac{[CiR]}{K_R}\right)^n}$$
where $[CiA],\ [CiR],\ K_A, \  K_R,\ a, \ r, \  \delta$ are positive real numbers,  $n\geq 1$ is a natural exponent   and $c$ is a real constant bigger or 
equal to $1$.
Then  $G$ is decreasing with respect to $\delta$ if and only if $[CiA]$ and $[CiR]$ verify
\begin{equation}\label{CondAttenuated}
a\frac{[CiA]}{\delta K_A}+r\frac{[CiR]}{\delta K_R } > \frac{[CiA]}{\delta K_A}+\frac{[CiR]}{\delta K_R } 
\mbox{ or equivalently }
G(\delta) > 1\, ,
\end{equation}
and increasing otherwise.
\end{lemma}

{\bf Proof.}
This assertion can be easily  verified computing
$$\frac{\partial G}{\partial \delta} = \frac{\left(\left(1-\frac1c \right) \frac{n\delta^{n-1}}{c} \big(h(\beta) - h( \alpha)\big) + \frac{n}{c^2}\alpha^{n-1} \beta^{n-1} (\beta -\alpha) \right) }{\Big(\left(1-\frac1c \right) \delta^n  +\frac1c \beta^n \Big)^{2}}$$
where $\alpha= \delta+ac\frac{[CiA]}{K_A}+rc\frac{[CiR]}{K_R}$, $\beta= \delta+c\frac{[CiA]}{K_A}+c\frac{[CiR]}{K_R}$ and $h(x) = x^{n-1} (x-\delta)$.
In the case of cooperativity ($c\geq 1$) the sign of this expression depends only on the differences $h(\beta) - h( \alpha)$ and $\beta -\alpha$.
Indeed both differences take always the same sign since $\alpha , \beta> \delta$ and $h(x)$ is an strictly increasing function for $x \geq \delta$.
Condition \eqref{CondAttenuated} is equivalent to $\beta < \alpha$ implying the negative character of  $\frac{\partial G}{\partial \delta}$ while
in the opposite case, $\beta > \alpha$,  $\frac{\partial G}{\partial \delta}$ is positive by the previous considerations. \hfill $\Box$

On the other hand, the regulation factor is monotonous increasing and decreasing with respect the total cooperation constant in the activated and repressed zones, i.e.,
\begin{lemma}
Let us consider the function
$$
H (c)  = 
\frac{\left(1-\frac1c\right)  +\frac1c \left(1+c a\frac{[CiA]}{K_A}+c r\frac{[CiR]}{\delta K_R }\right)^n}{\left(1-\frac1c\right) +\frac1c\left(1+c\frac{[CiA]}{K_A}+c\frac{[CiR]}{K_R}\right)^n}
= 
\frac{c-1   +\left(1+ c a\frac{[CiA]}{K_A}+c r\frac{[CiR]}{K_R }\right)^n}{c-1 +\left(1+ c \frac{[CiA]}{K_A}+ c \frac{[CiR]}{K_R}\right)^n}$$
where $[CiA],\ [CiR],\ K_A, \  K_R,\ a, \ r, $ are positive real numbers,  $n\geq 1$ is a natural exponent   and $c$ is a real constant bigger or 
equal to $1$.
Then  $H$ is increasing with respect to $c$ if and only if $[CiA]$ and $[CiR]$ verify
\begin{equation}\label{CondAttenuated2}
a\frac{[CiA]}{ K_A}+r\frac{[CiR]}{ K_R } > \frac{[CiA]}{ K_A}+\frac{[CiR]}{ K_R } 
\mbox{ or equivalently }
H(c) > 1\, ,
\end{equation}
and decreasing otherwise.
\end{lemma}

{\bf Proof.}
In this case the derivative of the functional $H$ can be expressed as
\begin{eqnarray}
\frac{\partial H}{\partial c} & =& 
\frac{\left( 1   +n \gamma \left(1+ c \gamma \right)^{n-1} \right) \Big(c-1 +\left(1+ c \epsilon \right)^n\Big)  }{\left(c-1  +\left(1+ c \epsilon \right)^n\right)^2}
  -
 \frac{\left( 1   +n \epsilon \left(1+ c \epsilon \right)^{n-1} \right) \Big(c-1 +\left(1+ c \gamma \right)^n\Big)  }{\left(c-1  +\left(1+ c \epsilon \right)^n\right)^2} 
  \nonumber \\
 & =  &
\frac{\left(c-1  +\left(1+ c \gamma \right)^n\right) }{\left(c-1  +\left(1+ c \epsilon \right)^n\right)}
\big( g(\gamma) - g(\epsilon) \big)
 \nonumber 
\end{eqnarray}
where $\gamma= a\frac{[CiA]}{K_A}+r\frac{[CiR]}{K_R}$, $\epsilon= \frac{[CiA]}{K_A}+\frac{[CiR]}{K_R}$.
This derivative is positive if and only if the inequality $g(\gamma) > g(\epsilon)$ holds being 
$g(x) = \frac{1   +n x \left(1+ c x \right)^{n-1}}{c-1 +\left(1+ c x \right)^n}$ a monotone increasing function. 
In consequence, $\frac{\partial H}{\partial c}$ will be positive if and only if $\gamma > \epsilon$ concluding the proof.  \hfill $\Box$

\subsection{Parameters}

For the sake of  biological reliability the values employed to generate blue solid lines in Figures \ref{FIG.displaced_threshold}\textbf{(b)},\textbf{(d)},\textbf{(f)} and \ref{FIG.afinidad_global_diferenciada_cooperatividad_global_diferenciada}, and the conserved TFs concentration \eqref{CiTot} in Figure \ref{FIG.displaced_threshold}\textbf{(a)},\textbf{(c)},\textbf{(e)},\textbf{(f)} have been adapted from fittings developed in\cite{Parker2011}:
\vspace{0.2cm} 

\begin{table}[htb]
\begin{center}
\begin{tabular}{|c|c||c|}
\hline
BEWARE constant & $C_B$ & $1\, nM\text{min}^{-1}$ \\
\hline
Activator transcription intensity & $a$ & $4.35$ \\
\hline
Repressor transcription intensity & $r$ & $5\times 10^{-5}$ \\
\hline
Activator binding affinity & $K_A$ & $9\times 10^{1}\, nM$ \\
\hline
Repressor binding affinity & $K_R$ & $3.1\times 10^{2}\, nM$ \\
\hline
RNA polymerase binding affinity & $K_{RP}$ & $[RNAP]$ \\
\hline
RNA polymerase concentration & $[RNAP]$ & $K_{RP}$ \\
\hline
Total cooperativity constant & $c$ & $1$  \\
\hline
Activator partial cooperativity constant & $c_A$ & $1$ \\
\hline
Repressor partial cooperativity constant & $c_R$ & $1$ \\
\hline
Cubitus total concentration & $h$ & $24\, nM$ \\
\hline
Cubitus gradient steepness &$D$ & $593$
\\
\hline
\end{tabular}
\end{center}
\caption{}
\label{tabla}
\end{table}


\end{document}